\documentclass[singlecolumn,12pt]{asme2ej}

\usepackage{epsfig} 
\usepackage{graphicx}
\usepackage[caption=false]{subfig}
\usepackage{amsmath}
\usepackage{float}
\usepackage{placeins}
\usepackage[justification=centering]{caption}
\title{Data-driven prediction of complex flow field over an axisymmetric body of revolution using Machine Learning}

\author{J. P. Panda\thanks{Address all correspondence to this author.}
    \affiliation{
    Department of Mechanical Engineering,\\
	DIT University, Dehradun, UK, India\\
    e-mail: jppanda@dituniversity.edu.in
    }
}
\author{H. V. Warrior
  \affiliation{Department of Ocean Engineering and Naval Architecture\\IIT Kharagpur, WB, India\\
 e-mail: warrior@naval.iitkgp.ac.in
     }	
 }
\begin{document}

\maketitle
\begin{abstract}
{\it Computationally efficient and accurate simulations of the flow over axisymmetric bodies of revolution (ABR) has been an important desideratum for engineering design. In this article the flow field over an ABR is predicted using machine learning (ML) algorithms, using trained ML models as surrogates for classical computational fluid dynamics (CFD) approaches. The flow field is approximated as functions of x and y coordinates of locations in the flow field and the velocity at the inlet of the computational domain. The data required for the development of the ML models were obtained from high fidelity Reynolds stress transport model (RSTM) based simulations. The optimal hyper-parameters of the trained ML models are determined using validation. The trained ML models can predict the flow field rapidly and exhibits orders of magnitude speed up over conventional CFD approaches. The predicted results of pressure, velocity and turbulence kinetic energy are compared with the baseline CFD data, it is found that the ML based surrogate model predictions are as accurate as CFD results. This investigation offers a framework for fast and accurate predictions for a flow scenario that is critically important in engineering design.       
}
\end{abstract}
\begin{nomenclature}
\entry{$U$}{Mean velocity}
\entry{$u$}{Velocity in x- direction}
\entry{$v$}{Velocity in y- direction}
\entry{$w$}{Velocity in z- direction}
\entry{$w$}{Inlet velocity}
\entry{$Re_v$}{Volumetric Reynolds number}
\entry{$R_{ij}$}{Reynolds stress tensor}
\entry{$b_{ij}$}{Reynolds stress anisotropy}
\entry{$\delta_{ij}$}{Kronecker delta}
\entry{$k$}{turbulent kinetic energy}
\entry{$\rho$}{density}
\entry{$\nu$}{Kinematic viscosity}
\end{nomenclature}

\section{Introduction}

The flow over axisymmetric bodies of revolution has been an important topic of study for decades due to their central importance in fields such as aerospace engineering, ocean and naval engineering, etc. A primary parameter in the design of LTA systems (Lighter-Than-Air) such as Airships and Aerostats involves determining the drag at various speeds over their aerodynamic envelopes that are very accurately represented as axisymmetric bodies of revolution in CFD simulations \cite{reddy2018cfd}. Novel space launch vehicle often use the hammerhead design that are approximated by ABRs in CFD simulations\cite{disotell2017design}. Similarly, ABR simulations are essential in the design of new autonomous underwater vehicle (AUV) hulls \cite{akolekar2020cfd}. While such ABR configurations have been the central focus of theoretical and empirical studies, such historical data is limited in spatial and temporal resolution. All these factors necessitate repeated CFD simulations for different ABR configurations across different fields of engineering design. This is a significant computational burden over numerous iterations involved in the engineering design process. As a practical step to reduce this burden many investigators utilize simpler eddy-viscosity based turbulence models. But these simpler models have severe limitations and can make flawed predictions for ABR simulations due to the flow separation and streamline curvature\cite{williams2020experimental,disotell2017design}. In this scenario a framework that can offer computationally inexpensive, robust and accurate predictions would be advantageous to these applications. 

While numerical simulations based on computational fluid dynamics methods are the most popular alternative for flow field prediction especially in complex engineering cases, the rapid evolving of machine learning and data storage capacities, researchers working in the field of fluid dynamics, have started applying ML principles for flow field modeling and prediction for two dimensional and low Reynolds number flows \cite{brunton2020machine}. More recently, researchers working in the field of turbulence modeling have applied ML concepts for developing turbulence models at different levels, starting from eddy viscosity models (EVM) to Reynolds stress models and large eddy simulations (LES). In EVM,  data-driven models were learned for components of Reynolds stresses from high fidelity data sets of direct numerical simulations (DNS) or experiments. In LES, models were proposed for the sub-grid scale stresses. \cite{ling2016reynolds} proposed data-driven models for the Reynolds stress anisotropy tensor by using deep neural networks (DNN). It was noticed that the data-driven model prediction of the flow field was better than the predictions of baseline eddy viscosity and nonlinear eddy viscosity models. \cite{panda2021modelling} used DNN to develop models for the pressure strain correlation of turbulence \cite{panda2019review}. The model was trained using high fidelity DNS data sets of turbulent channel flow at different Reynolds numbers. \cite{wang2018investigations} have proposed a model for the subgrid-scale stress in LES. It was observed that the ML-based surrogate model was more accurate than the Smagorinsky models for isotropic turbulent flow predictions. Few other popular nominations of application of ML in turbulence modeling are \cite{tracey2015machine,xie2020artificial,duraisamy2019turbulence}.




\begin{figure*}[ht]
\centering
\subfloat[]{\includegraphics[width=0.8\textwidth]{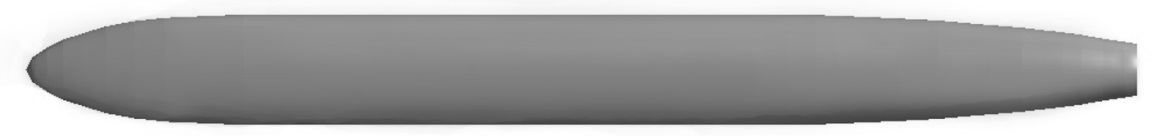}}\\
\subfloat[]{\includegraphics[width=0.8\textwidth]{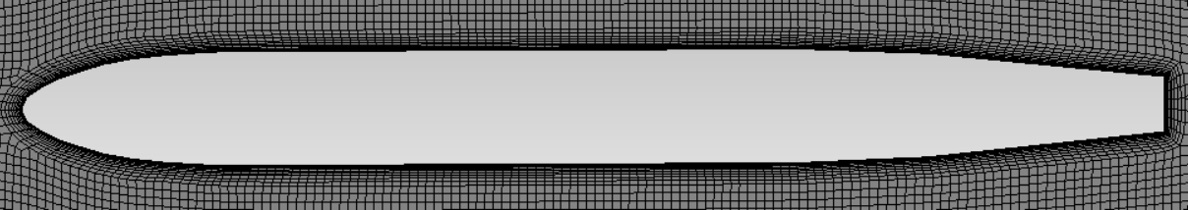}}
\caption {a) Geometry of the ABR. The length and diameter of ABR are 1.4 and 0.14 meters respectively, b) mesh for CFD simulations \label{fig:abr_geometry}}
\end{figure*}
In addition to the application of ML in turbulence modeling, several researchers have applied ML algorithms in the direct modeling of flow fields by correlating velocity and pressure with the input parameters or the boundary conditions. \cite{sekar2019fast} used DNN for prediction of flow field over airfoils. They used x, y, Reynolds number, angle of attack, and shape parameters as input data to the DNN model and outputs are pressure and velocity. x, y are the coordinates of different locations in the fluid domain. Numerical simulations data at different Reynolds numbers mainly in the laminar regime were performed to prepare a data-set for training and testing of the DNN flow prediction model. \cite{hui2020fast} used CNN for the prediction of the pressure distribution across an airfoil. The data set for the CNN model development was prepared using numerical simulations of deformed airfoils. \cite{renganathan2020machine} used DNN to develop surrogate models for transonic flow prediction past an airfoil. Using the DNN based model they predicted the flow field past a RAE2822 airfoil under varying freestream Mach numbers and angles of attack. Their main aim is to reduce the computational cost without compromising the accuracy of prediction. \cite{renganathan2020machine} utilized CNN to predict flow field in a scramjet isolator. Numerical simulation data at different Mach numbers and back pressures were utilized to develop the surrogate model. Using CNN they mainly mapped a relationship between the wall pressure and velocity field on the isolator. \cite{hasegawa2020machine} used CNN and LSTM for flow field prediction along bluff bodies. They used flow field data of 100 different bluff bodies for training and validation of the ML model. Out of 100 bluff body data, 80 data were used for training, and the rest 20 data were used to validate the ML model. \cite{lee2021analysis} predicted the unsteady volume wake flow field by using CNN. Past information of velocity and pressure were used to train and validate the model. Important work has also been done to estimate errors and uncertainties in the predictions of turbulence models using machine learning approaches. For example, \cite{heyse2021estimating} used an ensemble learning approach constrained by physics-based requirements to generate uncertainty bounds RANS model predictions. This investigation also utilized physics-based constraints to improve the ML model by imparting physics-based domain knowledge to it.
\cite{leer2021fast} predicted the flow field past bluff bodies using ANN. The data set for ANN model development was generated using CFD simulations. The ANN model was used to predict the unknown flow fields. \cite{kashefi2021point} utilized a novel deep learning architecture (point-net) for a one-to-one mapping of the flow field for flow past irregular geometries. The spatial position of the mesh nodes was taken as inputs to the DL model and corresponding velocity and pressure were taken as output. The non-linear relationship between the input and output was learned by the DL model. Numerical simulation data for laminar incomprehensible flow past a cylinder was used to train the model.

In this work, deep neural networks (DNN) and random forests (RF) were used to develop surrogate models for the data-driven prediction of the flow field of an ABR. Its shape is very much similar to the DARPA sub-off sub-marine model \cite{jagadeesh2009experimental, huang1978stern}. The data required to train and test the ML models were obtained from RSTM based simulations over the ABR. The x,y coordinates, and velocity at the inlet of the domain were considered as inputs of the ML model and the outputs to the ML model were velocity, pressure, and turbulence kinetic energy. The ML models have the capability to learn the non-linear relationship between the inputs and outputs. Most of the researchers have developed ML-based surrogate models for flow past airfoils at low-Reynolds numbers, but in this article, the turbulent flow field (U, P, and k) both in the boundary layer and wake region of the ABR is predicted using ML-based surrogate models, that is the novelty of this research.      
\section{ML Algorithms}
ML algorithms such as polynomial regression (PLR), support vector machines (SVR), random forests(RF), gradient boosted trees(GBT) and deep neural networks (DNN) can be used for flow field prediction problems since a function can be approximated by use of such techniques. In this article, we only have considered RF and DNN for the prediction of flow fields both in the boundary layer and wake region of the ABR.
\begin{figure*}
\centering
\includegraphics[height=6cm]{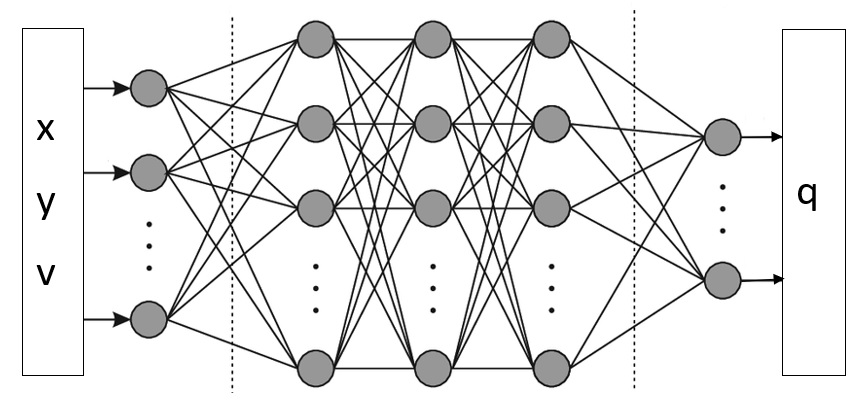}
\caption{The flow field prediction DNN. The optimized DNN has 5 layers with 120 neurons in each layers. In the diagram, x, y correspond to x and y-coordinates respectively and V is the velocity at the inlet of the computational domain. The arrow marks signifies the information flow from features to levels. q is the quantity of interest (u,p and k) that needs to be predicted.} \label{fig:3}
\end{figure*}

\subsection{Artificial Neural Networks (ANN)}
Artificial Neural Networks (ANN) (fig.\ref{fig:3}) are inspired by biological neural networks. These are machine learning systems and are also known as multi-layer perception (MLP). ANN with more than two hidden layers is known as deep neural networks (DNN). The function of biological neural networks (BNN) is to carry out a specific work when activated. The BNNs are the combination of neurons interconnected by synapses. In contrast to BNNs, ANN/MLP consists of artificial neurons.  A perceptron (fig.\ref{fig:3}a) is a basic unit of ANN. A non-linear activation function $\eta$ activates the linear combination of input data from the (l-1)th layer multiplied by a weight:    
\begin{equation}
\begin{aligned}
& q_{i}^{l}=\eta(\sum_j W_{ij}^{l}q_{j}^{l-1}).
\end{aligned}
\end{equation}
A number of perceptrons can be combined to form an ANN. A typical ANN is presented in fig.\ref{fig:3}b. The back-propagation algorithm can be used to calculate gradients for the weights of an ANN. The loss of the ML model can be minimized by using the gradient descent method.   
The activation functions, which can be used in MLP are sigmoid $\eta(\beta)=1/(1+e^{-\beta})$, hyperbolic tangent(tanh) $\eta(\beta)=(e^{-\beta}-e^{-\beta}).(e^{-\beta}+e^{-\beta})$ and RELU $\eta(\beta)=max[0,\beta]$.

\subsection{Random forests (RF)}
\begin{figure*}
\centering
\includegraphics[height=6.2cm]{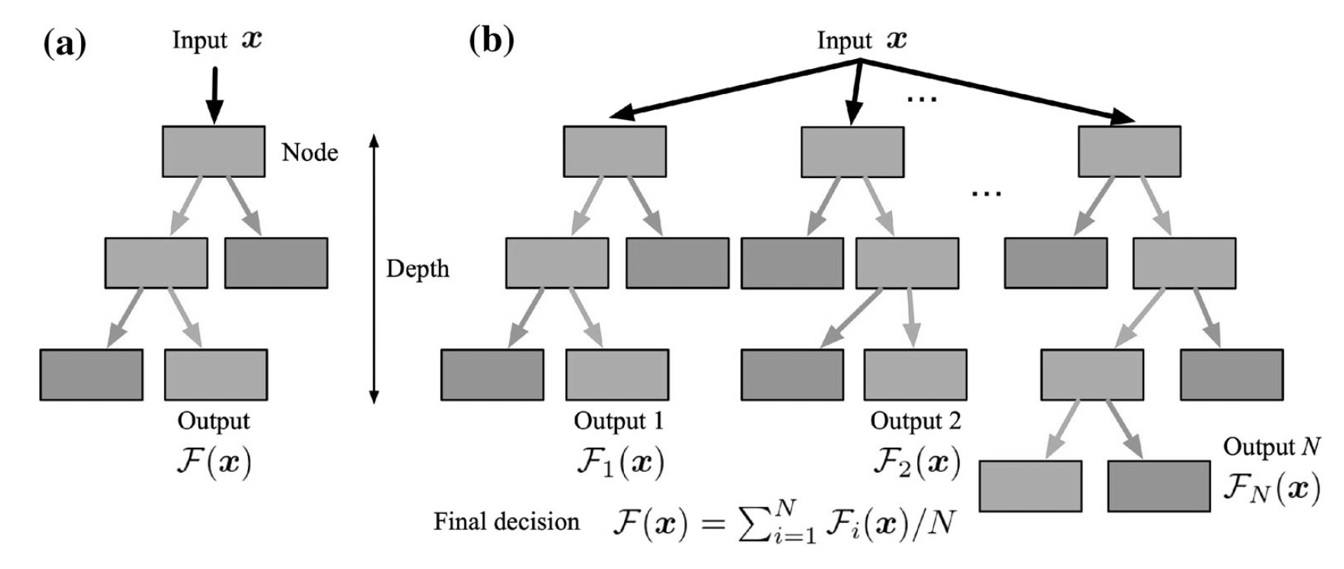}\\
\caption{Architecture of of a random forest, a) a single decision tree b) a random forest. The figure is reproduced from \cite{fukami2020assessment}} \label{fig:4}
\end{figure*}
\cite{breiman2001random} proposed the random forest algorithm. A decision tree (fig.\ref{fig:4}a) is the building block of an RF. The group of features and data are represented in boxes in the decision tree. The desired output can be obtained in the decision tree by applying. The output variable in an RF can be estimated by combining the performance of multiple regression trees (fig.\ref{fig:4}b). The RF is an assembled learning technique, which works on the bagging method concept. The regression trees in RF are prepared by utilizing a subset of random samples drawn with replacement from the data used in training. The trees are not allowed to prun in the RF and random vectors are generated to depict the growth of the trees. At each and every node of the RF, for the splitting of the data-set, a random combination of features is selected.

\section{Details of the Learning Dataset}
In data-driven machine learning applications, the learning dataset is centrally important as it determines the amount, quality, and nature of information available to the machine learning model framework. The learning dataset is traditionally split into a training dataset, used for optimization to tune the model coefficients; the validation dataset used to select hyperparameters of the model, and the testing dataset used to generate an unbiased estimate of the model's generalization error. In this investigation, we utilize a learning dataset generated using Reynolds Stress Transport Model (RSTM) based simulations. These are often used as surrogates for data from direct numerical studies or experimental investigations, especially in the generation of ML models. As opposed to eddy viscosity-based models where a constitutive relation is used to relate the turbulent quantities to the mean stresses, RSTM simulations utilize separate evolution equations for each component of the Reynolds stress tensor\cite{mishra2010pressure}. RSTM simulations account for the directional effects of turbulence evolution, the intercomponent energy transfer effected by the non-local action of pressure, and are able to deal with high degrees of anisotropy\cite{mishra2013intercomponent,panda2018representation}. RSTM simulations are also accurate for cases with high degrees of streamline curvature and flow separation as is observed in ABR simulations\cite{mishra2017toward,panda2021numerical}. The Reynolds stress transport equations can be written as follows:
\begin{equation}
\begin{split}
&\partial_{t} \overline{u_iu_j}+U_k \frac{\partial \overline{u_iu_j}}{\partial x_k}=P_{ij}-\frac{\partial T_{ijk}}{\partial x_k}-\eta_{ij}+\phi_{ij},\\
&\mbox{where},\\ 
& P_{ij}=-\overline{u_ku_j}\frac{\partial U_i}{\partial x_k}-\overline{u_iu_k}\frac{\partial U_j}{\partial x_k},\\
&\ T_{ijk}=\overline{u_iu_ju_k}-\nu \frac{\partial \overline{u_iu_j}}{\partial{x_k}}+\delta_{jk}\overline{ u_i \frac{p}{\rho}}+\delta_{ik}\overline{ u_j \frac{p}{\rho}},\\
&\epsilon_{ij}=-2\nu\overline{\frac{\partial u_i}{\partial x_k}\frac{\partial u_j}{\partial x_k}}  \\
&\phi_{ij}= \overline{\frac{p}{\rho}(\frac{\partial u_i}{\partial x_j}+\frac{\partial u_j}{\partial x_i})}\\
\end{split}
\end{equation}
$P_{ij}$ denotes the production of turbulence, $T_{ijk}$ is the diffusive transport, $\epsilon_{ij}$ is the dissipation rate tensor and $\phi_{ij}$ is the pressure strain correlation. The pressure fluctuations are governed by a Poisson equation:
\begin{equation}
\frac{1}{\rho}{\nabla}^2(p)=-2\frac{\partial{U}_j}{\partial{x}_i}\frac{\partial{u}_i}{\partial{x}_j}-\frac{\partial^2 u_iu_j}{\partial x_i \partial x_j}
\end{equation}
The fluctuating pressure term can be separated as rapid and slow pressure term $p=p^S+p^R$. The fluctuations in rapid and slow pressure satisfies the following equations.
\begin{equation}
\frac{1}{\rho}{\nabla}^2(p^S)=-\frac{\partial^2}{\partial x_i \partial x_j}{(u_iu_j-\overline {u_iu_j})}
\end{equation}
\begin{equation}
\frac{1}{\rho}{\nabla}^2(p^R)=-2\frac{\partial{U}_j}{\partial{x}_i}\frac{\partial{u}_i}{\partial{x}_j}
\end{equation}
The slow and rapid pressure terms account for the turbulence-turbulence and turbulence-mean strain interactions respectively. The rational mechanics approach was used to model the pressure strain correlation \cite{pope2001turbulent}

\begin{equation}
\phi_{ij}^R=4k\frac{\partial{U}_l}{\partial{x_k}}(M_{kjil}+M_{ikjl})
\end{equation}
where, 
\begin{equation}
M_{ijpq}=\frac{-1}{8\pi k}\int \frac{1}{r} \frac {\partial^2 R_{ij}(r)}{\partial r_p \partial r_p}dr
\end{equation}
where, $R_{ij}(r)=\langle u_i(x)u_j(x+r) \rangle$
For homogeneous turbulence the complete pressure strain correlation can be written as
\begin{equation}
\phi_{ij}=\epsilon A_{ij}(b)+kM_{ijkl}(b)\frac{\partial\overline {v}_k}{\partial{x_l}}
\end{equation}
In this work, we have used the quadratic Reynolds stress model (SSG pressure strain correlation model \cite{speziale1991modelling}):  

\begin{equation}
\begin{split}
& \phi_{ij}= -(C_1 \rho \epsilon + C_1^* P)  b_{ij} + C_2 \rho \epsilon (b_{ik}b_{kj}-\frac{1}{3} b_{mn}b_{mn}\delta_{ij}) + \\ &
(C_3-C_3^*\sqrt{b_{ij}b_{ij}}) \rho K S_{ij}+ C_4 \rho K(b_{ik} S_{jk}+b_{jk} S_{ik}-2/3b_{mn} S_{mn}\delta_{ij})\\ & +C_5 \rho K (b_{ik} W_{jk}+b_{jk} W_{ik})
\end{split}
\end{equation}

Here $b_{ij}=\frac{\overline{u_iu_j}}{2k}-\frac{\delta_{ij}}{3}$ is the Reynolds stress anisotropy tensor, $S_{ij}$ is the mean rate of strain and $W_{ij}$ is the mean rate of rotation. 

The numerical simulations were performed using the commercial CFD solver ANSYS Fluent \cite{ansys} at 11 different Reynolds numbers for the flow past ABR, where the in-compressible form of Navier-Stokes equations are solved. We have used SIMPLE(Semi-Implicit Method for Pressure
Linked Equations) scheme for the pressure velocity coupling. The mesh was generated in ANSYS meshing utility. Adequate y+ values were used for the RSTM model and inflation with a ratio of 1.2 was considered up to 10 layers in the boundary layer. The simulations were performed for 2-dimensional turbulent flow cases. The free stream velocities of water are 0.2, 0.4, 0.6, 0.8, 1, 1.2, 1.4, 1.6, 1.8, 2, 2.2 and 2.4 m/s respectively.  
\begin{figure}
\centering
\includegraphics[height=4cm]{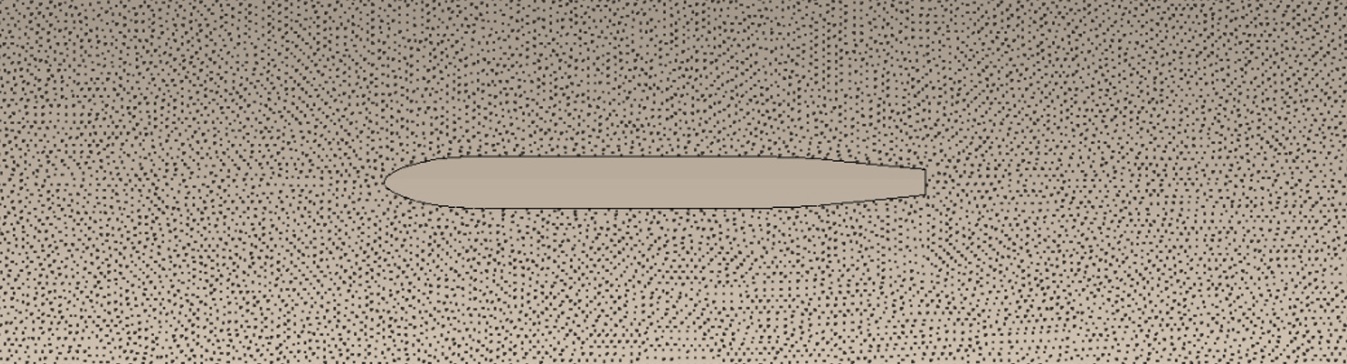}
\caption{Point cloud across the ABR.} \label{fig:pointcloud}
\end{figure}

\section{Data preparation for ML model development}
Numerical results were collected at 10000 different locations across the AUV to form the train-test data-set. We have considered a sub-domain (x=-1~2.4 and y=-0.5~0.5) across the ABR for data collection. The actual size of the CFD domain is, x=-1~5 and y=-1~1. At the nose tip of the ABR, x=0, and y=0. A point cloud (fig. \ref{fig:pointcloud}) was generated across the AUV in CFD post. For 11 Reynolds numbers, the total dataset size is 110000. We have made six different columns in the excel sheet, which are x, y, V, U, P, and k. x,y, and V are the input features, and U, P, k are the output levels respectively. V stands for free stream velocity, the rest of the symbols has their usual meaning. We have prepared two different training and testing data sets for the development of ML-based predictive models. Dataset1.0 has all other data except 1.0 m/s data as a training set and the data at 1.0 m/s is kept for testing the ML models. Similarly, dataset2.0 has all other data except 2.0 m/s data as a training set, and the data at 2.0 m/s is kept for testing the ML models. All the features in the datasets were normalized to a common scale using the min-max scaler of the SK-Learn \cite{scikit-learn}. The range of the features was taken as [-1,1]. Data normalization helps in making a common scale for the features and maintains the differences in a range of values, that preserves general distribution and ratios in the features of the dataset.        
\begin{figure}
\centering
 \subfloat[]{\includegraphics[height=7cm]{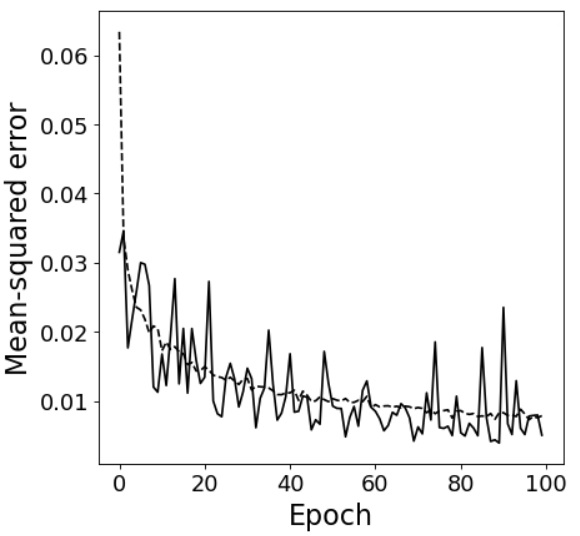}}\\
 \subfloat[]{\includegraphics[height=7cm]{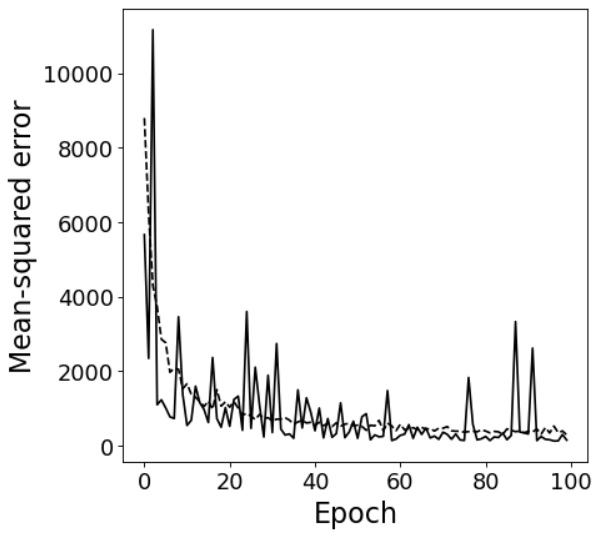}}\\
 \subfloat[]{\includegraphics[height=7cm]{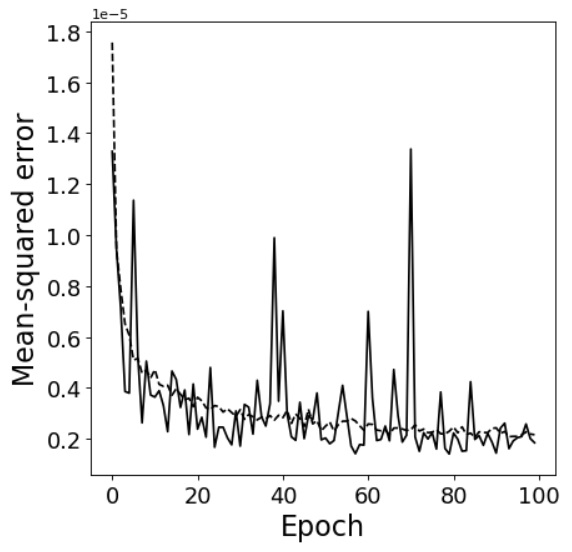}}
\caption{Loss history of the DNN training, the dotted and solid lines correspond to training and validation losses respectively. a) u, b) p and c) k losses.} \label{fig:loss}
\end{figure}
\section{ML model Training and validation}
For training and testing of ANN Tensorflow with Keras API \cite{abadi2016tensorflow} and for the random forest, SKLearn \cite{scikit-learn} was used. The ML models are utilized here to find a mapping function between the inputs and output quantities of interest. So the task accomplished is basically a regression.

The mapping function for both the ML algorithms can be written as: 
\begin{equation}
\begin{split}
(U, P, k) = f_{ML} (x, y, V) 
\end{split}
\end{equation}
where, U, P, k are the outputs and x, y and V are the inputs to the ML models. $f_{ML}$ is the approximation function. The neural networks were trained in such a manner that the final error of prediction is minimized. The loss vs epoch figures is shown in figure \ref{fig: loss}. Here mean squared error (MSE) was considered as loss for the training of the ANN. The MSE is mainly used to determine the performance of an algorithm, it can be defined as:

\begin{equation}
MSE=\frac{1}{N} \sum_{i=1}^{N}(\phi_i-\hat{\phi}_i)^2
\end{equation}

The training and test $R^2$ scores are presented in table \ref{t: Rsquare_nn} and \ref{t: Rsquare_rf}. The $R^2$ can be defined as:   

\begin{equation}
R^2=1- \frac {\sum_i(\phi_i-\hat{\phi}_i)^2} {\sum_i(\phi_i-\bar{\phi_i})^2}
\end{equation}

The neural networks are named as NNx, where x is the digit, that corresponds to number of layers in that neural network. Each layer in the NNx has 120 neurons. From the table, it is clear that there is a sharp increase in both train and test values of $R^2$, which signifies a better correlation between the input and output data. We did not increase the number of layers further to avoid over-fitting (over-fitting is a case that arises when with an increase in the size of neural network train $R^2$ increases, but there is the reduction of $R^2$ for the validation). The train and test $R^2$ for different random forests are shown in table \ref{t: Rsquare_rf}. The important hyper-parameters of random forests are the number of decision trees and their depth. From the table, it is noticed that the Random forest with 50 decision trees and a depth of 50 has the best value of $R^2$ for both training and testing.

\begin{table*}[h]
\begin{center}
\begin{tabular} {c c c c c c} 
  \bfseries  &  \multicolumn{2}{c}{} \\
  \bfseries Models &  \multicolumn{1}{c}{\bfseries NN1} & \multicolumn{1}{c}{\bfseries NN2} &
  \multicolumn{1}{c}{\bfseries NN3} &
  \multicolumn{1}{c}{\bfseries NN4} &
  \multicolumn{1}{c}{\bfseries NN5} 
  \\

Train $R^2$ & 0.948 & 0.991 & 0.988 & 0.991 & 0.988\\
Test $R^2$ & 0.949 & 0.849 & 0.936 & 0.946 & 0.966\\
\end{tabular}\\
\caption{$R^2$(coefficient of determination) of $v$ predictions by training-test cases by different NN models. The number beside NN signifies the number of layers in the NN. There are 120 neurons in each layer of the NN.}\label{t: Rsquare_nn}
\end{center}
\end{table*}

\begin{table*}[h]
\begin{center}
\begin{tabular} {c c c c c c} 
  \bfseries  &  \multicolumn{2}{c}{} \\
  \bfseries Models &  \multicolumn{1}{c}{\bfseries RF1} & \multicolumn{1}{c}{\bfseries RF2} &
  \multicolumn{1}{c}{\bfseries RF3} &
  \multicolumn{1}{c}{\bfseries RF4} &
  \multicolumn{1}{c}{\bfseries RF5} 
  \\

Train $R^2$ & 0.982 & 0.996 & 0.995 & 0.995 & 0.995\\
Test $R^2$ & 0.481 & 0.930 & 0.914 & 0.921 & 0.935\\
\end{tabular}\\
\caption{$R^2$(coefficient of determination) of $v$ predictions by training-test cases by different RF models. RF1, RF2, RF3, RF4 and RF5 has hyper-parameters (10,10), (20,20), (30,30), (40,40), (50,50) respectively. In the brackets the first and second number correspond to number of decision trees and their maximum depth respectively.}\label{t: Rsquare_rf}
\end{center}
\end{table*}
\begin{figure}
\centering
\subfloat[]{\includegraphics[height=7cm]{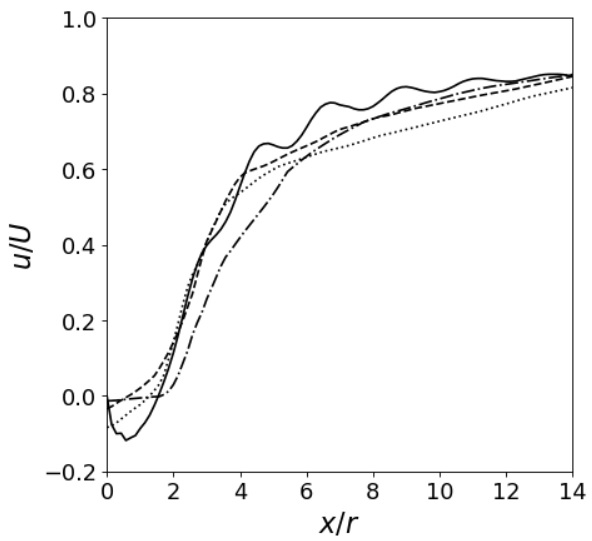}}\\
\subfloat[]{\includegraphics[height=7cm]{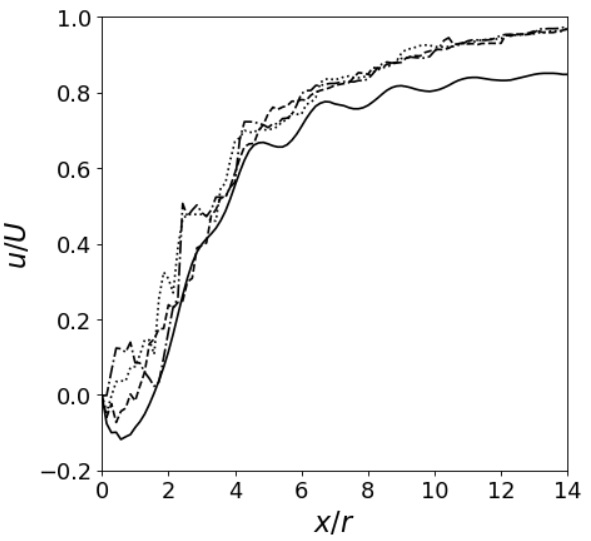}}
\caption{Effect of data size on the predictive capability of ML models} \label{fig:DSS}
\end{figure}

\subsection{Effect of data-set size on predictive capability of the ML models}  
In supervised learning, the dataset size can have a significant effect on the model training. Small training dataset results in a poor approximation function. In order to study the effect of dataset size on the predictive capability of the ML models, we have prepared three separate training datasets each having 30000, 50000, and 100000 data. The data was collected from CFD posts using appropriate point clouds. The dataset with 30000 samples has 3000 data for each of the Reynolds numbers. In a similar fashion, we have created the other two datasets. All the three datasets were used in the training of DNN and RF separately, with the optimized hyperparameters. This study may be named as Data Sensitivity Study (DSS), similar to the mesh sensitivity study performed in CFD simulations. The DSS is proposed in this paper for the first time. The DSS results are presented in the figure. \ref{fig:DSS}. The first and second figures correspond to the DNN and RF model perditions respectively. The predictions are for the test data of 1 m/s. In both, the figure's solid lines correspond to the baseline RSM results of the velocity field in the wake region of ABR at an inlet velocity of 1 m/s. The dashed, dotted and dashed-dot lines correspond to velocity predictions using ML models trained with 100000, 50000, and 30000 data respectively. Figure. \ref{fig:DSS}a and b correspond to DNN and RF model predictions respectively. From both the figures it is clear that with an increase in dataset size the predictive capability of both DNN and RF increases. We did not increase data size further to save computational time. We have performed DSS for the other two variables e.g pressure and turbulence kinetic energy, but for brevity, we have presented only the prediction results of the velocity field in the wake region of the ABR.   
\begin{figure}
\centering
\subfloat[]{\includegraphics[height=7cm]{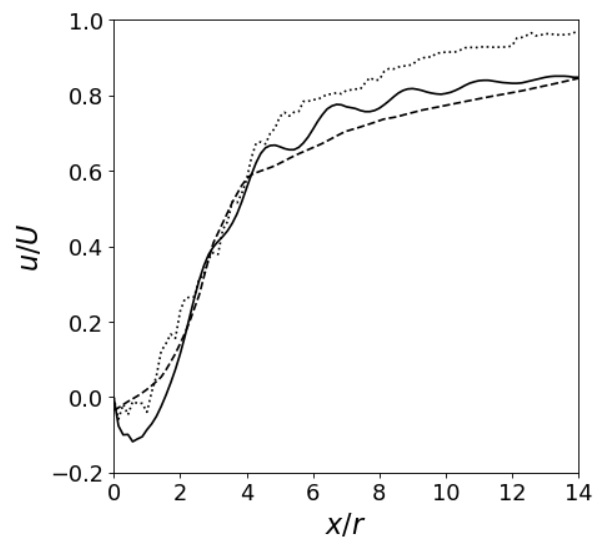}}\\
\subfloat[]{\includegraphics[height=7cm]{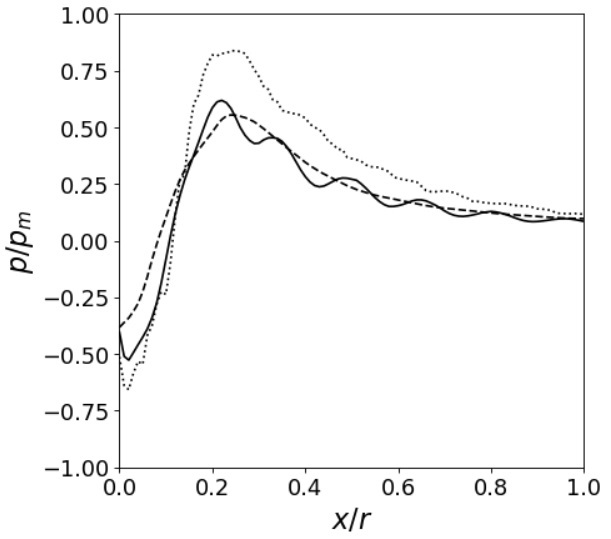}}\\
\subfloat[]{\includegraphics[height=7.1cm]{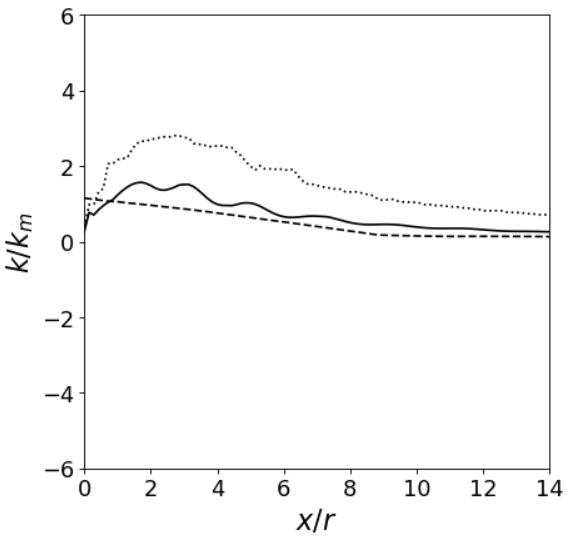}}
\caption{Profiles of velocity, pressure and turbulence kinetic energy in the wake region of the ABR for inlet velocity 1 m/s, Solid line RSM, dashed line neural network, dotted line random forest, y/r=0.} \label{fig:1ms}
\end{figure}
\begin{figure}
\centering
\includegraphics[height=7cm]{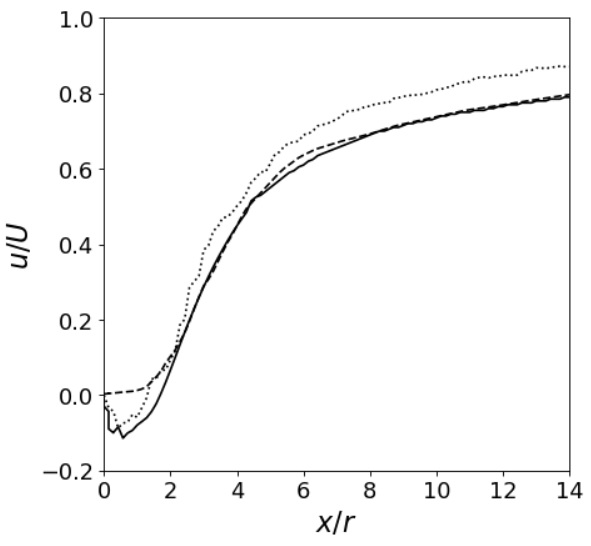}\\
\includegraphics[height=7cm]{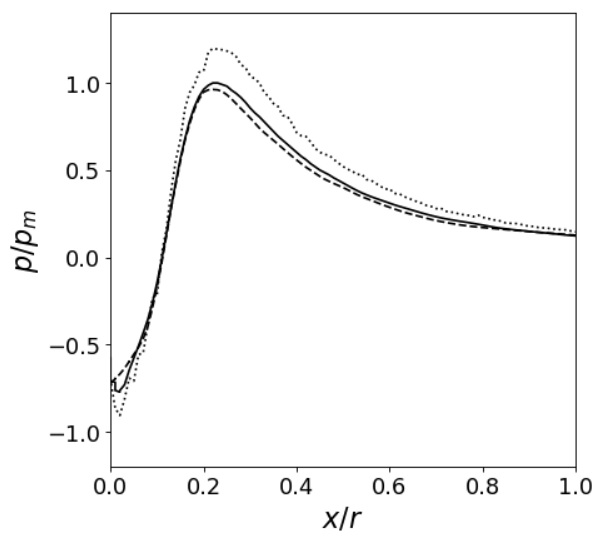}\\
\includegraphics[height=7.1cm]{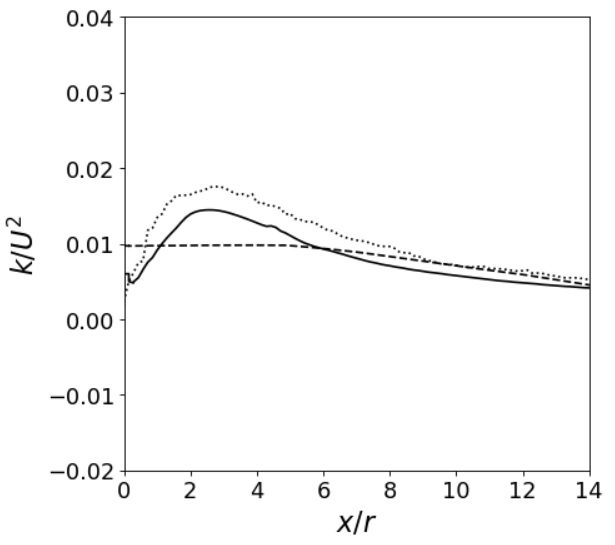}
\caption{Profiles of velocity, pressure and turbulence kinetic energy in the wake region of the ABR for inlet velocity 2 m/s, Solid line RSM, dashed line neural network, dotted line random forest, y/r=0.} \label{fig:2ms}
\end{figure}
\begin{figure}
\centering
\includegraphics[height=7cm]{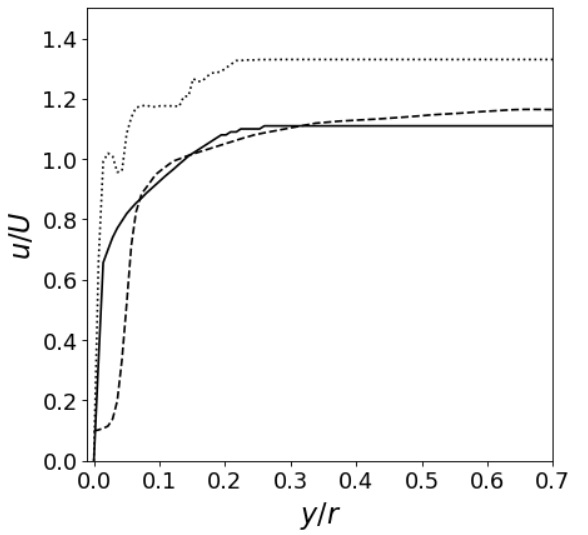}\\
\includegraphics[height=7cm]{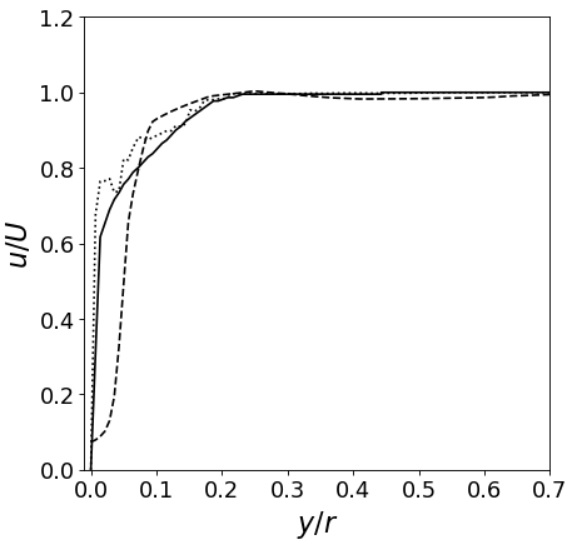}
\caption{Profiles of velocity in the boundary layer of the ABR for inlet velocity (a) 1 m/s and (b) 2 m/s, Solid line RSM, dashed line neural network, dotted line random forest, x/r=10.} \label{fig:boundaryvel}
\end{figure}

\section{Prediction of unknown flow field}
We have applied the optimized DNN and RF for the prediction of the unknown flow cases, those were not used for training the models. As discussed in the data preparation section, those are flowing past the ABR at 1 and 2 m/s respectively. In figures. \ref{fig:1ms}, the DNN and RF Predictions are presented by dashed and dotted lines respectively. The solids lines baseline results of RSM models. The DNN predictions of the velocity, pressure, and turbulent kinetic energy match well with RSM results. The RF prediction of the velocity field is as good as the RSM results but the TKE and pressure predictions poor in contrast to the DNN predictions. Similar is the case for flow predictions at 2 m/s as shown in figure \ref{fig:2ms}. The figures \ref{fig:1ms} and \ref{fig:2ms} correspond to the distribution of velocity, pressure, and turbulence kinetic energy in the wake region of the ABR. In addition to wakefield prediction, the flow field in the boundary layer of the ABR was also presented in the figure \ref{fig:boundaryvel}. From both the figures, it is clear that the DNN prediction of the flow field is better than the predictions of the RF model. This concludes that the DNNs can act as a universal approximator in predicting the flow field of the ABR. Hence DNN based surrogate data-driven ML models can be applied in the prediction of the flow field in the preliminary design stages of the ABR.

\section{Conclusions}
In this article, we introduce a framework that can offer computationally inexpensive, robust, and accurate predictions for flow simulations over Axisymmetric Bodies of Revolution. We have applied different ML algorithms such as artificial neural networks and random forests for flow field prediction around an ABR. The flow statistics around the ABR were utilized to develop the ML models. Mainly, regression functions for the velocity, pressure and turbulence kinetic energy were approximated using the input velocity and x,y coordinates as inputs to the ML models. The optimal hyper-parameters of the ML models were presented. It is noticed that the ML predicted flow fields are as accurate as of the actual flow field as obtained by the RSM model. The ML models presented in this article will be useful for fast flow field predictions along the ABR in its preliminary design stage. In the future course of work, the Reynolds numbers and shape parameters can be incorporated in the modeling basis for improved prediction of the flow field around the ABR.

          
\newpage
\clearpage
\bibliographystyle{asmems4}
\bibliography{asme2e}
\clearpage
\end{document}